\documentclass[runningheads,amssymbols]{llncs}
\usepackage{multirow}

\title{Comparison between CPBPV, ESC/Java, CBMC, Blast, EUREKA and Why \\
for Bounded Program Verification}

\titlerunning{Comparison of program verification frameworks}
 
\author{H\'el\`ene Collavizza\inst{1},   Michel Rueher\inst{1}, Pascal  Van Hentenryck\inst{2}}

\institute{
  Universit\'e de Nice--Sophia Antipolis, France (\email{\{helen,rueher\}@polytech.unice.fr})\and
  Brown University, Box 1910, Providence, RI 02912 (\email{pvh@cs.brown.edu})
}

\begin{document}

\maketitle
\section{Abstract}
This  report describes experimental results for a set of benchmarks on
program verification.  It compares the capabilities of CPBVP {\em
  ``Constraint Programming framework  for Bounded Program
  Verification''} \cite{CRVH08} with  the following frameworks: ESC/Java,  CBMC, Blast, EUREKA and Why.

\section{Introduction}

 This  report describes experimental results for a set of benchmarks on  program verification.  It compares the capabilities of CPBVP {\em ``Constraint Programming framework  for Bounded Program Verification''} \cite{CRVH08} with the following frameworks: 

\begin{itemize}
\item ESC/Java (http://kind.ucd.ie/products/opensource/ESCJava2/):   Extended 
Static Checker for Java is a programming tool that attempts to find common
 run-time errors in JML-annotated Java programs by static analysis of the program code
 and its formal annotations. 

\item CBMC (http://www.cprover.org/cbmc/): is a Bounded Model Checker for ANSI-C and C++ programs. It allows verifying array bounds (buffer overflows), pointer safety, exceptions and user-specified assertions. 

\item Blast(http://mtc.epfl.ch/software-tools/blast/): Berkeley Lazy Abstraction Software Verification Tool is a software model checker for C programs.  

\item EUREKA (http://www.ai-lab.it/eureka/):  is a C bounded model checker which 
uses an SMT solver instead of an SAT solver.

\item Why (http://why.lri.fr/): is a software verification platform which  integrates
  many existing provers (proof assistants such as Coq, PVS, HOL 4,... and decision procedures such as Simplify, Yices, ...).
\end{itemize}

All experiments were performed on the same machine, 
an Intel(R) Pentium(R) M processor 1.86GHz with 1.5G of memory,
  using the version of the verifiers that can be downloaded from their web sites (except for EUREKA project, for which we report the execution times given by the authors in \cite{ABM07} and \cite{AMP06}).

For each benchmark program, we describe the data entries and the verification parameters. Since
 the input formats slightly differ from one framework to another,  we also give the input files
 that were used to perform the comparisons for each benchmark and each framework.
 In experimental result tables, 
UNABLE means that the framework is unable to validate the program (either because a lack of expression power or  time overflow), NOT\_FOUND   that it doesn't detect an error that was inserted in the program, and FALSE\_ERROR  that it finds an arror in a correct program.

\section{Triangle classification}

The {\em tritype} program is a standard benchmark in test case generation
and program verification since it contains numerous non-feasible
paths: only 10 paths correspond to actual inputs because of complex
conditional statements in the program. The program takes three
positive integers as inputs (the triangle sides) and returns 2 if the
inputs correspond to an isoscele triangle, 3 if they correspond to an
equilateral triangle, 1 if they correspond to some other triangle, and
4 otherwise (see \ref{tritype}).

\subsection{Program used for CPBPV, ESC/Java and Why}\label{tritype}

{\scriptsize
\begin{verbatim}
/** Triangle classification
 * returns 4 if (i,j,k) are not the sides of a triangle
 *   3 if (i,j,k) is an equilateral triangle
 *   2 if (i,j,k) is an isoscele triangle
 *   1 if (i,jk) is a scalene triangle
 **/

   /*@ requires (i >= 0 && j >= 0 && k >= 0);
     @ ensures
     @   (((i+j) <= k || (j+k) <= i || (i+k) <= j) ==> (\result == 4))
     @  && ((!((i+j) <= k || (j+k) <= i || (i+k) <= j) && (i==j && j==k)) ==> (\result == 3))
     @  && ((!((i+j) <= k || (j+k) <= i || (i+k) <= j) && !(i==j && j==k) && (i==j || j==k || i==k)) ==> (\result == 2))
     @  && ((!((i+j) <= k || (j+k) <= i || (i+k) <= j) && !(i==j && j==k) && !(i==j || j==k || i==k)) ==> (\result == 1));
     @*/
1   int tritype (int i, int j, int k) {
2      int trityp;
3      if (i == 0 || j == 0 || k == 0) {
4        trityp = 4;}
5     else {
6       trityp = 0;
7       if (i == j) {trityp = trityp + 1;}
8       if (i == k) {trityp = trityp + 2;}
9       if (j == k) {trityp = trityp + 3;}
10      if (trityp == 0) {
11         if ((i+j) <= k || (j+k) <= i || (i+k) <= j) {
12             trityp = 4;}
13         else {trityp = 1;}         
14      }
15      else {
16        if (trityp > 3) {trityp = 3;}
17        else {
18         if (trityp == 1 && (i+j) > k) {
19            trityp = 2;}
20          else {
21            if (trityp == 2 && (i+k) > j) {
22               trityp = 2;}
23            else {
24              if (trityp == 3 && (j+k) > i) {
25                 trityp = 2;}
26              else {
27                trityp = 4;}
              }
            }
          }
        }
      }
      return trityp;   
   }
\end{verbatim}
}

\subsection{C program used for CBMC}\label{tritypeCBMC}

The only difference with the Java version is that we translated the ``implies'' statement of JML specification with the corresponding disjunction (ie $a \Rightarrow b$ is translated as $\neg a \vee b$).

{\scriptsize
\begin{verbatim}
 int tritype(int i, int j, int k) {
    // PRECONDITION
    __CPROVER_assume(i>=0&&j>=0&&k>=0);
    int trityp ;
    if ( (i <= 0) || (j <= 0) || (k <= 0)){
      trityp = 4 ;
    }
    else {
        trityp = 0 ;
        if (i == j) trityp = trityp + 1 ;
        if (i == k) trityp = trityp + 2 ;
        if (j == k) trityp = trityp + 3 ;
        if (trityp == 0){
            if ( (i+j <= k) || (j+k <= i) || (i+k <= j)) {
              trityp = 4 ;
          }
            else{
              trityp = 1 ;
          }
        }
        else {
            if (trityp > 3) {
              trityp = 3 ;
            }
            else
              if ( (trityp == 1) && (i+j > k) ){
                 trityp = 2 ;
             }
              else
                if ( (trityp == 2) && (i+k > j) ){//ERROR trityp==1
                   trityp = 2 ;
              }
                else
                  if ( (trityp == 3) && (j+k > i)) {
                    trityp  = 2 ;
   	         }
                  else {
                    trityp = 4 ;
                  }
           }
      }
    // POSTCONDITION
    assert((!((i+j<=k)||(j+k<=i)||(i+k<=j)) || trityp == 4) &&   
      (!(!((i+j<=k)||(j+k<=i)||(i+k<=j))&&((i==j)&&(j==k)))
       || trityp == 3) &&                          
      (!(!((i+j<=k)||(j+k<=i)||(i+k<=j))&&!((i==j)&&(j==k))   
        &&((i==j)||(j==k)||(i==k))) || trityp == 2) &&   
      (!(!((i+j<=k)||(j+k<=i)||(i+k<=j))&&!((i==j)&&(j==k))   
        && !((i==j)||(j==k)||(i==k))) || trityp == 1));

    return trityp ;
  }
\end{verbatim}
}

\subsection{C program used for Blast}\label{tritypeBlast}

Blast is unable to deal with arithmetic expressions like {\em i+k$<$=j}
unless these expressions have been collected as  decisions  taken 
in a program path. For example, assertion in line 22 (see program below), can be verified because 
it directly results from the ``if'' statement on line 15. But assertion in line 34 can't be 
verified because it requires a reasoning on arithmetic expressions.
So we used a slighty different version of the tritype program for Blast.

{\scriptsize
\begin{verbatim}
#include <assert.h>

int main(int i, int j, int k)  {
1    int trityp ;
2    if ((i <= 0) || (j <= 0) || (k <= 0)){
3       trityp = 4 ;
4       assert((i <= 0) || (j <= 0) || (k <= 0));
5    }
6    else      {
7        trityp = 0 ;
8        if (i == j)
9           trityp = trityp + 1 ;
10       if (i == k) 
11          trityp = trityp + 2 ;
12       if (j == k )
13          trityp = trityp + 3 ;
14       if (trityp == 0)  {
15          if ((i+j <= k) || (j+k <= i) || (i+k <= j)) {
16              trityp = 4 ;
17              assert((i+j<=k)||(j+k<=i)||(i+k<=j));
18          }
19          else{
20              trityp = 1 ;
21              assert((i!=j) && (j!=k) && (i!=k)
22                      && !((i+j<=k)||(j+k<=i)||(i+k<=j))  );
23          }
24       }
25       else {
26            if (trityp > 3) {
27              trityp = 3 ;
28              assert((i==j && j==k && i==k));
29            }
30            else
31              if ((trityp == 1) && (i+j > k) ){
32                 trityp = 2 ;
33                 assert(i==j );
34                 //assert(!((i+j<=k)||(j+k<=i)||(i+k<=j)));
35              }
36              else
37                if ((trityp == 2) && (i+k > j) ){ //ERROR trityp==1
38                   trityp = 2 ;
39                   assert(i==k );
40                }
41                else
42                  if ((trityp == 3) && (j+k > i)) {
43                     trityp  = 2 ;
44                     assert(j==k);
45                  }
46                  else {
47                     trityp = 4 ;
48                     assert((i+j<=k)||(j+k<=i)||(i+k<=j));
                    }
           }
      }
    return trityp ;
  }
\end{verbatim}}

\subsection{Comparative results}

Table \ref{tabTritype} shows experimental results for {\em Tritype} program
using CPBPV, ESC/Java, CBMC, BLAST  and Why frameworks. Note that BLAST was unable to validate this example because the current version does not handle  linear arithmetic. But it succeeded in verifying the 
easier version presented in section \ref{tritypeBlast} in $0.716s$.

Note that our previous
approach  using constraint programming and Boolean abstraction to abstract the conditions,
validated this benchmark in $8.52$ seconds  when
integers were coded on 16 bits \cite{CoR06}. It also explored 92
spurious paths.

\begin{table}[t]
\begin{small}
\begin{center}
\begin{tabular}{|c|c|c|c|c|c||c|}
 \hline
    & CPBPV & ESC/Java & CBMC   & Why &  BLAST & BLAST (easier version)\\
\hline
time & 0.287s & 1.828s &  0.82s  &  8.85s &UNABLE &0.716s \\
 \hline
\end{tabular}
\end{center}
\caption{Comparison table for Tritype program}\label{tabTritype}

\end{small}
\end{table}

\section{Triangle classification with an error}

In this section, we consider an erroneous version of {\em Tritype} program where we have replaced the 
test {\em ``if ((trityp==2)\&\&(i+k$>$j))''} in line 22 (see section \ref{tritype}) with the test {\em ``if ((trityp==1)\&\&(i+k$>$j))''}.

 Since the local variable 
{\em trityp}  is equal to {\em 2} when {\em i==k},  if {\em (i+k)$>$j}  we know that  {\em (i,j,k)} are the sides of an 
isoscele triangle. In fact, the two other triangular inequalities $i+j>k$ and $j+k>i$ 
are trivial because j$>$0.
But when {\em trityp=1}, {\em i==j} and this erroneous version can answer that
 the triangle is isoscele while it may  not be a triangle at all (the triangular inequality
$i+j>k$ or $j+k>i$ may not be verified).
For example, it will return {\em 2} when {\em (i,j,k)=(1,1,2)}.

\subsection{Program used for CPBPV, ESC/Java and Why}

We show below the programs used for CPBPV, ESC/Java and Why.
The program for Blast was modified in a similar way.

{\scriptsize
\begin{verbatim}
 /* an error has been inserted line 21: trityp==1 instead of 2*/

   /*@ requires (i >= 0 && j >= 0 && k >= 0);
     @ ensures
     @   (((i+j) <= k || (j+k) <= i || (i+k) <= j) ==> (\result == 4))
     @  && ((!((i+j) <= k || (j+k) <= i || (i+k) <= j) && (i==j && j==k)) ==> (\result == 3))
     @  && ((!((i+j) <= k || (j+k) <= i || (i+k) <= j) && !(i==j && j==k) && (i==j || j==k || i==k)) ==> (\result == 2))
     @  && ((!((i+j) <= k || (j+k) <= i || (i+k) <= j) && !(i==j && j==k) && !(i==j || j==k || i==k)) ==> (\result == 1));
     @*/

1   int tritypeKO (int i, int j, int k) {
2      int trityp;
3      if (i == 0 || j == 0 || k == 0) {
4        trityp = 4;}
5     else {
6       trityp = 0;
7       if (i == j) {trityp = trityp + 1;}
8       if (i == k) {trityp = trityp + 2;}
9       if (j == k) {trityp = trityp + 3;}
10      if (trityp == 0) {
11         if ((i+j) <= k || (j+k) <= i || (i+k) <= j) {
12             trityp = 4;}
13         else {trityp = 1;}         
14      }
15      else {
16        if (trityp > 3) {trityp = 3;}
17        else {
18         if (trityp == 1 && (i+j) > k) {
19            trityp = 2;}
20          else {
21            if (trityp == 1 && (i+k) > j) { //ERROR: trityp==1 instead of 2
22               trityp = 2;}
23            else {
24              if (trityp == 3 && (j+k) > i) {
25                 trityp = 2;}
26              else {
27                trityp = 4;}
              }
            }
          }
        }
      }
      return trityp;   
   }
\end{verbatim}}

\subsection{Program used for CBMC}\label{tritypeKOCBMC}

We show below the program used for CBMC. The main function was used to
run the C program in order to verify that the program contains an error.

{\scriptsize
\begin{verbatim}
#include <assert.h>
#include <stdio.h>

 int tritype(unsigned int i,unsigned int j,unsigned int k) {
    int trityp ;
    if ( (i <= 0) || (j <= 0) || (k <= 0)){
      trityp = 4 ;
    }
    else   {
        trityp = 0 ;
        if ( i == j)
          trityp = trityp + 1 ;
        if ( i == k) 
          trityp = trityp + 2 ;
        if ( j == k )
          trityp = trityp + 3 ;
        if (trityp == 0)  {
            if ((i+j <= k) || (j+k <= i) || (i+k <= j)) {
              trityp = 4 ;
            }
            else { trityp = 1 ;  }
        }
        else  {
            if (trityp > 3) {
               trityp = 3 ; }
            else
              if ((trityp == 1) && (i+j > k)){
                 trityp = 2 ; }
              else
                if ((trityp == 1) && (i+k > j)){ // ERROR: trityp == 1 instead of 2
                   trityp = 2 ; }
                else
                  if ((trityp == 3) && (j+k > i)) {
                    trityp  = 2 ; }
                  else {
                    trityp = 4 ; }
          }
      }
    assert((!((i+j<=k)||(j+k<=i)||(i+k<=j)) || trityp == 4) &&   
	   (!(!((i+j<=k)||(j+k<=i)||(i+k<=j))&&((i==j)&&(j==k)))
              || trityp == 3) &&                          
	   (!(!((i+j<=k)||(j+k<=i)||(i+k<=j))&&!((i==j)&&(j==k))   
              &&((i==j)||(j==k)||(i==k))) || trityp == 2) &&   
	   (!(!((i+j<=k)||(j+k<=i)||(i+k<=j))&&!((i==j)&&(j==k))   
              && !((i==j)||(j==k)||(i==k))) || trityp == 1));
    return trityp ;
  }
      
int main(void) {
  int t = tritype(1,1,2);
  printf("trityp %i\n",t);
  return 0;
}
\end{verbatim}}

\subsection{Comparative results}

Table \ref{tabTritypeKO} shows experimental results for the {\em erroneous} version of {\em Tritype} program 
for CPBPV, ESC/Java, CBMC, BLAST  and Why. Execution times correspond to the time required to find the first error. 

For frameworks that were able to find the error, we give in section \ref{tritypeTrace}
the error traces printed by the framework.

\paragraph{\bf Remark on results with CBMC}

Note that for CBMC framework, CBMC is unable to detect the error but
 when  running  the C  program for values $(i,j,k)=(1,1,2)$, the assertion 
verification mechanism of C detects that the assertion is violated.

If we use ``CPROVER\_assert'' instead of ``assert'' (as recommended by D. Kroening when we have
contacted him),
then CBMC finds the error in the erroneous version of tritype.
Nevertheless, if we also use this option in the correct version of the tritype program,
then CBMC finds a false error. The reason seems to be that CBMC works using modulo arithmetic
and so we must specify that there is no 
overflow. So, we also added the statement:
\begin{quotation}
 $CPROVER_assume(i+j>=0\&\&j+k>=0\&\&k+i>=0)'$
\end{quotation}

\noindent which means that there is no overflow intohe sums.

\begin{table}[t]
\begin{small}
\begin{center}
\begin{tabular}{|c|c|c|c|c|c||c|}
 \hline
    & CPBPV & ESC/Java & CBMC   & WHY & BLAST & BLAST (easier version)\\
\hline
time &  0.056s s & 1.853s & NOT\_FOUND  & NOT\_FOUND & UNABLE & 0.452s \\
 \hline
\end{tabular}
\end{center}
\caption{Comparison table for Tritype program with error}\label{tabTritypeKO}
\end{small}
\end{table}

\subsection{Error traces}\label{tritypeTrace}

We give here the execution traces of the three frameworks that were able to find the error.

\paragraph{\bf CPBPV error trace}
{\scriptsize 
\begin{verbatim}
i_0[-2147483647:2147483646] : 1
j_0[-2147483647:2147483646] : 1
k_0[-2147483647:2147483646] : 2
trityp_0[-2147483647:2147483646] : 0
trityp_1[-2147483647:2147483646] : 0
trityp_2[-2147483647:2147483646] : 1
trityp_3[-2147483647:2147483646] : 2
\end{verbatim}}

The result is variable {\em trityp\_3} which is equal to {\em 2}.
The two sides {\em i} and {\em j} are equals but {\em (i,j,k)}
doesn't represent a triangle because the triangular inequality is not verified
(i.e {\em i+j=k}). So returned value must be {\em 4} (part 1 of the JML specification).

\paragraph{\bf ESC/Java error trace}
{\scriptsize
\begin{verbatim}
TritypeKO.java:67: Warning: Postcondition possibly not established (Post)
        }
        ^
Associated declaration is "TritypeKO.java", line 12, col 5:
          @ ensures ...
            ^
Execution trace information:
    Executed else branch in "TritypeKO.java", line 23, col 7.
    Executed then branch in "TritypeKO.java", line 25, col 15.
    Executed else branch in "TritypeKO.java", line 28, col 3.
    Executed else branch in "TritypeKO.java", line 31, col 3.
    Executed else branch in "TritypeKO.java", line 42, col 8.
    Executed else branch in "TritypeKO.java", line 46, col 9.
    Executed else branch in "TritypeKO.java", line 50, col 10.
    Executed then branch in "TritypeKO.java", line 51, col 39.
    Executed return in "TritypeKO.java", line 66, col 2.

Counterexample context:
    (0 < k:18.32)
    ((2 * j:18.25) <= k:18.32)
    (k:18.32 <= intLast)
    (longFirst < intFirst)
    (1000001 <= intLast)
    (null <= max(LS))
    (eClosedTime(elems) < alloc)
    (vAllocTime(this) < alloc)
    ((intFirst + 1000001) <= 0)
    (intLast < longLast)
    (0 <= j:18.25)
    (k:18.32 == 0) == tmp0!cor:20.6
    null.LS == @true
    (null <= max(LS))
    typeof(j:18.25) <: T_int
    ((j:18.25 + k:18.32) > j:18.25) == @true
    (0 + 1) == 1
    (j:18.25 == 0) == tmp1!cor:20.6
    typeof(k:18.32) <: T_int
    typeof(this) <: T_TritypeKO
    ((j:18.25 + j:18.25) > k:18.32) == tmp4!cand:47.9
    typeof(this) <: T_TritypeKO
    trityp:19.6<7> == 2
    T_bigint == T_long
    tmp0!cor:20.23 == tmp0!cor:20.6
    trityp:19.6<2> == 1
    trityp:19.6<5> == 2
    elems@pre == elems
    j:18.25 == i:18.18
    trityp:19.6<8> == 2
    tmp5!cand:51.25 == @true
    trityp:19.6 == 2
    trityp:26.4 == 1
    trityp:19.6<3> == 1
    state@pre == state
    trityp:19.6<6> == 2
    tmp1!cor:20.13 == tmp1!cor:20.6
    trityp:19.6<1> == 1
    tmp5!cand:51.13 == @true
    alloc@pre == alloc
    tmp4!cand:47.21 == tmp4!cand:47.9
    !typeof(this) <: T_void
    !T_java.lang.Object <: T_java.io.Serializable
    typeof(this) != T_void
    bool$false != @true
    tmp4!cand:47.9 != @true
    ecThrow != ecReturn
    1 != 0
    k:18.32 != j:18.25
    k:18.32 != 0
    this != null
    trityp:19.6<7> != 4
    tmp0!cor:20.23 != @true
    j:18.25 != 0
    tmp1!cor:20.6 != @true
\end{verbatim}}

\paragraph{\bf CBMC trace}

{\scriptsize
\begin{verbatim}
Counterexample:
State 15 file bsearchAssertKO.c line 10 function binsearch thread 0
----------------------------------------------------
  bsearchAssertKO::binsearch::1::low=0 (00000000000000000000000000000000)
State 16 file bsearchAssertKO.c line 10 function binsearch thread 0
----------------------------------------------------
  bsearchAssertKO::binsearch::1::high=7 (00000000000000000000000000000111)
State 17 file bsearchAssertKO.c line 11 function binsearch thread 0
----------------------------------------------------
  bsearchAssertKO::binsearch::1::result=-1 (11111111111111111111111111111111)
State 18 file bsearchAssertKO.c line 13 function binsearch thread 0
----------------------------------------------------
  bsearchAssertKO::binsearch::1::1::middle=3 (00000000000000000000000000000011)
State 21 file bsearchAssertKO.c line 17 function binsearch thread 0
----------------------------------------------------
  bsearchAssertKO::binsearch::1::high=2 (00000000000000000000000000000010)
State 25 file bsearchAssertKO.c line 13 function binsearch thread 0
----------------------------------------------------
  bsearchAssertKO::binsearch::1::1::middle=1 (00000000000000000000000000000001)
State 29 file bsearchAssertKO.c line 15 function binsearch thread 0
----------------------------------------------------
  bsearchAssertKO::binsearch::1::high=0 (00000000000000000000000000000000)
State 33 file bsearchAssertKO.c line 13 function binsearch thread 0
----------------------------------------------------
  bsearchAssertKO::binsearch::1::1::middle=0 (00000000000000000000000000000000)
State 37 file bsearchAssertKO.c line 15 function binsearch thread 0
----------------------------------------------------
  bsearchAssertKO::binsearch::1::high=-1 (11111111111111111111111111111111)
Violated property:
  file bsearchAssertKO.c line 21 function binsearch
  assertion
  result != -1 && a[result] == x || result == -1 && a[0] != x && a[1] != x 
      && a[2] != x && a[3] != x && a[4] != x && a[5] != x && a[6] != x && a[7] != x
VERIFICATION FAILED
\end{verbatim}}

\paragraph{\bf Blast trace}
{\scriptsize
\begin{verbatim}
Start trace
       XXX
0 :: 0:          FunctionCall(__BLAST_initialize_tritypeKO.i()) :: -1
      XXX
0 :: 0:          Block(Return(0);) :: -1
      XXX
-1 :: -1:        Skip :: 27
      XXX
27 :: 27:        Pred(i@main  >  0) :: -1
      XXX
27 :: 27:        Pred(j@main  >  0) :: -1
      XXX
27 :: 27:        Pred(k@main  >  0) :: -1
      XXX
33 :: 33:        Block(trityp@main = 0;) :: 34
      XXX
34 :: 34:        Pred(i@main  ==  j@main) :: -1
      XXX
35 :: 35:        Block(trityp@main = trityp@main  +  1;) :: 36
      XXX
36 :: 36:        Pred(i@main  !=  k@main) :: -1
      XXX
38 :: 38:        Pred(j@main  !=  k@main) :: -1
      XXX
40 :: 40:        Pred(trityp@main  !=  0) :: -1
      XXX
54 :: 54:        Pred(trityp@main  <=  3) :: -1
      XXX
59 :: 59:        Pred(trityp@main  ==  1) :: -1
      XXX
59 :: 59:        Pred(i@main  +  j@main  <=  k@main) :: -1
      XXX
65 :: 65:        Pred(trityp@main  ==  1) :: -1
      XXX
65 :: 65:        Pred(i@main  +  k@main  >  j@main) :: -1
      XXX
66 :: 66:        Block(trityp@main = 2;) :: 67
      XXX
67 :: 67:        Pred(i@main  !=  k@main) :: -1
      XXX
67 :: 67:        FunctionCall(__assert_fail(__assertion@__assert_fail = "i==k",__file@__assert_fail = "tritypeKO.c",__line@__assert_fail = 67,__function@__assert_fail = "main",)) :: -1
      XXX
77 :: 77:        FunctionCall(__blast_assert()) :: -1
      XXX
End trace
\end{verbatim}
}

\section{Binary search}

In this section we consider the usual binary search program which determines
 if a value $x$ is present in a sorted
array $tab$ (see \ref{bsearchJava} for a Java version of this program).

\subsection{Java program used for CPBPV and ESC/Java}\label{bsearchJava}
{\scriptsize
\begin{verbatim}
  /*@ requires (\forall int i; (i >= 0 && i < tab.length -1); tab[i] <= tab[i+1]);
    @ ensures
    @  ((\result == -1) ==> (\forall int i; (i >= 0 && i < tab.length); tab[i] != x)) 
    @   && ((\result != -1) ==> (tab[\result] == x));
    @*/
  int binarySearch (int[] tab, int x) {
    int index = -1;
    int m = 0;
    int l = 0;
    int u = tab.length -1;
    while (index == -1 && l <= u) {
      m = (l + u) / 2;
      if (tab[m] == x) {
        index = m;      
      }
      else {
        if (tab[m] > x) {
          u = m - 1;        
        }
        else {
          l = m + 1;
        }
      }
    }
    return index;
  }
}
\end{verbatim}}

\subsection{C program for an instance of length 8 used with CBMC}

In order to express the {\em forall}  statements of the JML specification  inside the CBMC framework, 
 we unfolded the conditions for  fixed array lengths. The program below shows the preconditions and 
postconditions for an array of length 8.
We proceeded in the same way for other array lengths.

{\scriptsize
\begin{verbatim}
int binsearch(int x) { 
  int a[8];
  // PRECONDITION
  __CPROVER_assume(a[0]<=a[1]&&a[1]<=a[2]&&a[2]<=a[3]&&a[3]<=a[4]
       &&a[4]<=a[5]&&a[5]<=a[6]&&a[6]<=a[7]);

  signed low=0, high=7;
  int result=-1;
  while(result==-1&&low<=high) { 
    signed middle=(high+low)/2;
    if(a[middle]<x)
      high=middle-1;
    else if(a[middle]>x)
      low=middle+1;
    else // a[middle]=x !
      result= middle;
  }
  // POSTCONDITION
  assert((result!=-1 && a[result]==x)||(result==-1 && (a[0]!=x&&a[1]!=x&&a[2]!=x&&
          a[3]!=x&&a[4]!=x&&a[5]!=x&&a[6]!=x&&a[7]!=x)));
  return result;
}
\end{verbatim}}

\subsection{Program with invariant used with Why}\label{whyBsearch}

This version of the binary search is given as example in the Why distribution.
It uses a loop invariant which allows Why to use induction when generating 
proof obligations. 

{\scriptsize
\begin{verbatim}
/*@ axiom mean_1 : \forall int x, int y; x <= y => x <= (x+y)/2 <= y */

/* binary_search(t,n,v) search for element v in array t
   between index 0 and n-1
   array t is assumed sorted in increasing order
   returns an index i between 0 and n-1 where t[i] equals v,
   or -1 if no element of t is equal to v
 */

/*@ requires
  @   n >= 0 && \valid_range(t,0,n-1) &&
  @   \forall int k1, int k2; 0 <= k1 <= k2 <= n-1 => t[k1] <= t[k2]
  @ ensures
  @   (\result >= 0 && t[\result] == v) ||
  @   (\result == -1 && \forall int k; 0 <= k < n => t[k] != v)
  @*/
int binary_search(int* t, int n, int v) {
  int l = 0, u = n-1;
  /*@ invariant
    @   0 <= l && u <= n-1 &&
    @   \forall int k; 0 <= k < n => t[k] == v => l <= k <= u
    @ variant u-l
    @*/
  while (l <= u ) {
    int m = (l + u) / 2;
    if (t[m] < v) l = m + 1;
    else if (t[m] > v) u = m - 1;
    else return m;
  }
  return -1;
}
\end{verbatim}}

\subsection{Comparative results}



Table \ref{tabsearch} reports comparative results for the binary search.

For  ESC/Java framework, the number of loop unfolding must be given.
Since the worst case complexity of binary search algorithm is 
$O(log(n))$ where $n$ is the array length,  we set the parameter ``Loop'' to $log(n)+1$.

In a similar way, within the CBMC framework, an overestimate of the  number of loop unfoldings 
is required (parameter ``unwind'').

Note that CPBPV doesn't require any additional information (neither invariant nor loop unfolding bound) because
at any time  the entrance condition of the loop is known. When performing symbolic execution, 
it  selects a path, taking decisions for conditional expressions  as ``if (tab[m]==x)''. 
These decisions  involve
 that the lower and upper bounds $l$ and $u$ are assigned with constant values.

\begin{table}[t]
\begin{small}
\begin{center}
\begin{tabular}{{|c||l|c|c|c|c|c|c|}}
 \hline
\multirow{2}*{CPBPV} & array length & 8 & 16 & 32 & 64 & 128 & 256 \\
\cline{2-8}
        & time & 1.081s & 1.69s & 4.043s & 17.009s & 136.80s& 1731.696s \\
 \hline
\multirow{2}*{CBMC} & array length & 8 & 16 & 32 & 64 & 128 & 256 \\
\cline{2-8}
        & time & 1.37s & 1.43s & TIMEOUT ($>$6000s)  & TIMEOUT &TIMEOUT &TIMEOUT \\
 \hline
\multirow{2}*{Why} & with invariant  & \multicolumn{6}{|l|}{11.18s} \\
\cline{2-8}
                   & without invariant & \multicolumn{6}{|l|}{UNABLE} \\
 \hline
ESC/Java & \multicolumn{7}{|l|}{FALSE\_ERROR} \\
 \hline
BLAST & \multicolumn{7}{|l|}{UNABLE}\\
 \hline
\end{tabular}
\end{center}
\caption {Comparison table for binary search}\label{tabsearch}
\vspace{-0.5cm}
\end{small}
\end{table}

The Why framework was very efficient to make the verification when an invariant is given as shown in subsection \ref{whyBsearch} but was unable to make it if no invariant is provided.

The CBMC framework was not able to do the verification for an instance of array
of length 32 (it was interrupted after 6691,87s).

 ESC/Java  found a false error in this program. 
 David Cok, a developper of ESC/Java we have contacted, has answered that 
 we need to add some loop invariants in order to be able to perform the proof.

\section{Binary search with error}

We  consider here an erroneous version of the binary search algorithm.
We update the lower bound and the upper bound in the same way,  whether the middle value is greater or less than the searched value (see line 15 in program below). We modified in the same way the binary search versions for CBMC and Why.

{\scriptsize
\begin{verbatim}
class BsearchKO {
  /*@ requires (\forall int i; (i >= 0 && i < tab.length -1); tab[i] <= tab[i+1]);
    @ ensures
    @  ((\result == -1) ==> (\forall int i; (i >= 0 && i < tab.length); tab[i] != x)) 
    @   && ((\result != -1) ==> (tab[\result] == x));
    @*/
  int binarySearch (int[] tab, int x) {
1    int index = -1;
2    int m = 0;
3    int l = 0;
4    int u = tab.length -1;
5    while (index == -1 && l <= u) {
6      m = (l + u) / 2;
7      if (tab[m] == x) {
8        index = m;      
9      }
10      else {
11        if (tab[m] > x) {
12          u = m - 1;        
13        }
14        else {
15          u = m - 1; //ERROR: u = m - 1 instead of l = m + 1;
        }
      }
    }
    return index;
  }
}
\end{verbatim}}

\subsection{Comparative results}

Table \ref{tabsearchKO} shows experimental results for {\em binary search} program
with {\em error} for CPBPV, ESC/Java, CBMC, and Why using an invariant. 

The Why framework was unable to perform this proof because  60\% of the proof obligations
remained unknown. 


\begin{table}[t]
\begin{small}
\begin{center}
\begin{tabular}{|c|c|c|c|c|c|}
 \hline
    & CPBPV & ESC/Java & CBMC   & WHY with invariant\\
\hline
  length  8 & 0.027s & 1.21 s  & unwind=4  1.38s  & NOT\_FOUND \\
\hline
 length  16 & 0.037s & 1.347 s  & unwind=6 1.69s  & NOT\_FOUND\\
\hline
 length  32 &  0.064s  &  1.792 s & unwind=7  7.62s  & NOT\_FOUND\\
\hline
 length  64 &  0.115s & 1.886 s  & unwind=8  27.05s  & NOT\_FOUND\\
\hline
 length  128 & 0.241s & 1.964 s  & unwind=9  189.20s  & NOT\_FOUND\\
 \hline
\end{tabular}
\end{center}

\caption{Comparison table for binary search with error}\label{tabsearchKO}
\end{small}
\end{table}

\subsection{Error traces}
We display here the error trace found with CPBPV for an array of length 8 and integers coded on 32 bits.

\paragraph{\bf CPBPV error trace}
{\scriptsize
\begin{verbatim}
Counter-example found
x_0[-2147483647:2147483646] : -2147483646
i_0[-2147483647:2147483646] : [-2147483647..2147483647]
i_0[-2147483647:2147483646] : [-2147483647..2147483647]
i_0[-2147483647:2147483646] : [-2147483647..2147483647]
i_0[-2147483647:2147483646] : [-2147483647..2147483647]
result_0[-2147483647:2147483646] : -1
milieu_0[-2147483647:2147483646] : 0
gauche_0[-2147483647:2147483646] : 0
droite_0[-2147483647:2147483646] : 7
milieu_1[-2147483647:2147483646] : 3
droite_1[-2147483647:2147483646] : 2
milieu_2[-2147483647:2147483646] : 1
droite_2[-2147483647:2147483646] : 0
milieu_3[-2147483647:2147483646] : 0
droite_3[-2147483647:2147483646] : -1
JMLResult_0[-2147483647:2147483646] : -1
tab_0[0][-2147483647:2147483646] : -2147483647
tab_0[1][-2147483647:2147483646] : -2147483647
tab_0[2][-2147483647:2147483646] : -2147483646
tab_0[3][-2147483647:2147483646] : -2147483645
tab_0[4][-2147483647:2147483646] : -2147483645
tab_0[5][-2147483647:2147483646] : -2147483645
tab_0[6][-2147483647:2147483646] : -2147483645
tab_0[7][-2147483647:2147483646] : -2147483645
\end{verbatim}}

\paragraph{\bf ESC/Java error trace}
We display here the error trace found with ESC/Java for all the possible array lengths.
Command line is:  escj -Loop 64.5 BsearchKO.java
{\scriptsize
\begin{verbatim}
BsearchKO.java:32: Warning: Postcondition possibly not established (Post)
        }
        ^
Associated declaration is "BsearchKO.java", line 8, col 5:
          @ ensures ...
            ^
Execution trace information:
    Reached top of loop after 0 iterations in "BsearchKO.java", line 17, col 2.
    Executed else branch in "BsearchKO.java", line 22, col 8.
    Executed else branch in "BsearchKO.java", line 26, col 9.
    Reached top of loop after 1 iteration in "BsearchKO.java", line 17, col 2.
    Executed return in "BsearchKO.java", line 31, col 2.
\end{verbatim}}

\paragraph{\bf CBMC error trace}
We display here the error trace found with CBMC for an array of length 
8 and parameter unwind sets to 6.

{\scriptsize
\begin{verbatim}
Counterexample:

State 1 file /usr/include/getopt.h line 59 thread 0
----------------------------------------------------
  optarg=NULL

State 2 file /usr/include/getopt.h line 59 thread 0
----------------------------------------------------
  optarg#str=NULL

State 3 file /usr/include/getopt.h line 73 thread 0
----------------------------------------------------
  optind=0 (00000000000000000000000000000000)

State 4 file /usr/include/getopt.h line 78 thread 0
----------------------------------------------------
  opterr=0 (00000000000000000000000000000000)

State 5 file /usr/include/getopt.h line 82 thread 0
----------------------------------------------------
  optopt=0 (00000000000000000000000000000000)

State 6 file /usr/include/stdio.h line 142 thread 0
----------------------------------------------------
  stdin=NULL

State 7 file /usr/include/stdio.h line 143 thread 0
----------------------------------------------------
  stdout=NULL

State 8 file /usr/include/stdio.h line 144 thread 0
----------------------------------------------------
  stderr=NULL

State 9 file <built-in> line 12 thread 0
----------------------------------------------------
  __CPROVER_alloc=(assignment removed)

State 10 file <built-in> line 13 thread 0
----------------------------------------------------
  __CPROVER_alloc_size=(assignment removed)

State 11 file /usr/include/bits/sys_errlist.h line 27 thread 0
----------------------------------------------------
  sys_nerr=0 (00000000000000000000000000000000)

State 12 file /usr/include/unistd.h line 474 thread 0
----------------------------------------------------
  __environ=NULL

State 15 file bsearchAssertKO.c line 10 function binsearch thread 0
----------------------------------------------------
  bsearchAssertKO::binsearch::1::low=0 (00000000000000000000000000000000)

State 16 file bsearchAssertKO.c line 10 function binsearch thread 0
----------------------------------------------------
  bsearchAssertKO::binsearch::1::high=7 (00000000000000000000000000000111)

State 17 file bsearchAssertKO.c line 11 function binsearch thread 0
----------------------------------------------------
  bsearchAssertKO::binsearch::1::result=-1 (11111111111111111111111111111111)

State 18 file bsearchAssertKO.c line 13 function binsearch thread 0
----------------------------------------------------
  bsearchAssertKO::binsearch::1::1::middle=3 (00000000000000000000000000000011)

State 21 file bsearchAssertKO.c line 17 function binsearch thread 0
----------------------------------------------------
  bsearchAssertKO::binsearch::1::high=2 (00000000000000000000000000000010)

State 25 file bsearchAssertKO.c line 13 function binsearch thread 0
----------------------------------------------------
  bsearchAssertKO::binsearch::1::1::middle=1 (00000000000000000000000000000001)

State 29 file bsearchAssertKO.c line 15 function binsearch thread 0
----------------------------------------------------
  bsearchAssertKO::binsearch::1::high=0 (00000000000000000000000000000000)

State 33 file bsearchAssertKO.c line 13 function binsearch thread 0
----------------------------------------------------
  bsearchAssertKO::binsearch::1::1::middle=0 (00000000000000000000000000000000)

State 37 file bsearchAssertKO.c line 15 function binsearch thread 0
----------------------------------------------------
  bsearchAssertKO::binsearch::1::high=-1 (11111111111111111111111111111111)

Violated property:
  file bsearchAssertKO.c line 21 function binsearch
  assertion
  result != -1 && a[result] == x || result == -1 && a[0] != x && a[1] != x && a[2] != x && a[3] != x && a[4] != x && a[5] != x && a[6] != x && a[7] != x

VERIFICATION FAILED
\end{verbatim}}

\section{Buble sort with initial condition}

This example is taken from
\cite{ABM07} and performs a bubble sort of an array $t$ which contains
integers from $0$ to $t.length$ given in decreasing order. The
$EUREKA$ tool \cite{ABM07} validates the benchmark for arrays of
lengths up to 8. In particular, it takes 91 seconds to verify for
length 8.

\subsection{Java program used for CPBPV and ESC/Java}\label{bubleJava}

{\scriptsize
\begin{verbatim}
/* Example taken from  Mantovani et all [SPIN'2006]
 * buble sort with a precondition
 */

class BubleSortMantovani {
  /*  @ requires (\forall int i; 0<= i && i < tab.length; tab[i] = tab.length -1-i);
      @ ensures (\forall int i; 0<= i && i < tab.length-1; tab[i]<=tab[i+1]);
  */
  void tri(int[] tab) {
     int i=0;
     while (i<tab.length-1){
        int j=0;
        while (j < tab.length-i-1) {
           if (tab[j]>tab[j+1]) {
              int aux = tab[j];
              tab[j]= tab[j+1];
              tab[j+1] = aux;
           }
           j++;
        }
        i++;
     }
   }
}
\end{verbatim}}

\subsection{C program for an instance of length 8 used for CBMC}\label{bubleCBMC}

{\scriptsize
\begin{verbatim}
 void buble() {
  int a[8];
  // PRECOND
  __CPROVER_assume(a[0]==7 &&a[1]==6 &&a[2]==5 &&a[3]==4 &&a[4]==3 &&a[5]==2 &&a[6]==1 &&a[7]==0 );

     int i=0;
     while (i<7){
        int j=0;
        while (j < 7-i) {
           if (a[j]>a[j+1]) {
              int aux = a[j];
              a[j]= a[j+1];
              a[j+1] = aux;
           }
           j++;
        }
        i++;
     }
     // POSTCONDITION
     assert(a[0]<=a[1]&&a[1]<=a[2]&&a[2]<=a[3]&&a[3]<=a[4]&&a[4]<=a[5]&&a[5]<=a[6]
           &&a[6]<=a[7]);
  }
\end{verbatim}}

\subsection{Comparative results}

Table \ref{tabbuble} shows the experimental results for the buble sort.

For the CPBPV framework, {\em  UNABLE} corresponds to a memory capacity overflow. This is due to 
the need of SSA-like array renaming to express successive assignments. 
In this first prototype, we did
 not carefully  manage  the memory and so we duplicated indexes of the array which have not changed. This could  easily be improved in a next version.

For ESC/Java framework, {\em  UNABLE} corresponds to the  message ``Caution: Unable to check method tri(int[]) of type BubleSortMantovani because its VC is too large''.

\begin{table}[t]
\begin{small}
\begin{center}
\begin{tabular}{|c|c|c|c|c|}
 \hline
    & CPBPV & ESC/Java & CBMC  & EUREKA\\
\hline
  length  8 & 0.031s & 3.778 s & 1.11s  &  91s\\
\hline
 length  16 & 0.032s & UNABLE &  2.01s  & UNABLE\\
\hline
 length  32 &  UNABLE  & UNABLE &  6.10s  & UNABLE\\
 \hline
 length  64 &  UNABLE  & UNABLE & 37.65s  & UNABLE\\
 \hline
\end{tabular}
\end{center}

\caption{Comparison table for buble sort}\label{tabbuble}
\end{small}

\end{table}

\section{Sum of the square of the n first integers}

This program computes the sum of the squares of the $n$ first integers.
The specification is that the sum is equal to $n\times(n+1)\times(n\times 2+1)/6$.
The main interest of this example is that it contains a non linear expression.

We didn't perform the verification with EUREKA and BLAST, because they do not deal with
non-linear expressions.

ESC/Java found a false error.

CBMC  was able to verify this program only if we add a precondition which set  $n$ to 
a constant value.

Table \ref{sumSquare} displays  comparative results.

\subsection{Java program used for CPBPV}

{\scriptsize
\begin{verbatim}
/** sum of the square of the n fisrt integers
 */
class SquareSum {

   /*@ requires (n >= 0); 
     @ ensures \result == (n*(n+1)*((n*2)+1))/6;
     @*/
   int somme (int n) {
      int i;
      int s = 0;
      while (i<=n) {
         s = s+i*i;
         i = i+1;   
      }
      return s;
   }
}
\end{verbatim}}

\subsection{C program used for CBMC}

In order to be able to perform the proof, we had to insert
a precondition which fixes the value of parameter n.

{\scriptsize
\begin{verbatim}
    int somme (int n) {
        // PRECONDITION
        __CPROVER_assume(n==8);
        int i=0;
        int s = 0;
        while (i<=n) {
            s = s+i*i;
            i = i+1;    
        }
        //POSTCONDITION
        assert(s==n*(n+1)*((n*2)+1)/6);
        return s;
    }
\end{verbatim}}

\begin{table}[t]
\begin{small}
\begin{center}
\begin{tabular}{|c|c|c|c|}
 \hline
    & CPBPV   & CBMC & ESC/Java\\
\hline
  length  8 &   0.152s  & 0.83s & FALSE\_ERROR\\
\hline
 length  16 &  0.557s & 0.85s & FALSE\_ERROR\\
\hline
 length  32 &   1.111s & 0.95s & FALSE\_ERROR\\
\hline
 length  64 &   1.144s & 1.13s & FALSE\_ERROR\\
\hline
 length  128 &    1.868s & 1.60s & FALSE\_ERROR\\
 \hline
\end{tabular}
\end{center}

\caption{Comparison table for sum of squares}\label{sumSquare}
\end{small}

\end{table}

\section{Sum of the square of any permutation of the n first integers}

This  benchmark illustrates some capabilities of CPBPV framework that are not handled
by other frameworks.
It emphasizes the ability
of specifying combinatorial constraints and of solving nonlinear
problems. The \verb+alldifferent+ constraint\cite{Reg94} in the
pre-condition specifies that all the elements of the array are
different, while the program constraints and postcondition involves
quadratic and cubic constraints.

This program takes two parameters as inputs: an array and its length. 
The array  contains any permutaiton of the integers  from $0$ to $n$. It returns the sum of the squares of 
the array elements, which must be equal to $n\times(n+1)\times(2\times n+1)/6$.

{\scriptsize
\begin{verbatim}
/**  Sum of the square of the n first integers
 * array t contains values between 0 and t.length-1 which are all different
 * (i.e array t contains any permutation of (0..t.length-1)
*/
class SquareSumArray {
/*
  @ requires (n == t.length-1) && 
  @    (\forall int i; 0<=i && i<t.length-1;0<=t[i]&&t[i]<=n) &&
  @    \alldifferent t;  // More compact notation than the JML  quantified formulae
  @ ensures \result == n*(n+1)*(2*n+1)/6;
  @*/
1 int sum(int[] t, int n) {
2    int s = 0;
3    int i = 0;
4    while (i!=t.length) {
5        s=s+t[i]*t[i]
6        i =i+1;    }
7    return s;}
\end{verbatim}}

\subsection{Experimental results}

The maximum instance that we were able to solve with CPBPV framework was an array of size 10
 in 66.179s.

\section{ Selection Sort}

This last benchmark  highlights both modular
verification and the {\tt element} constraint of constraint
programming to index arrays with arbitrary expressions. 

\subsection{Selection sort for modular verification}

{\scriptsize
\begin{verbatim}
/*@ ensures  (\forall int i; 0<=i && i<t.length-1;t[i]<=t[i+1]) @*/
1  static void selectionSort(int[] t) {
2    for (int i=0; i<t.length;i++){
3        int k = findMin(t,i);
5        int tmp = t[i];
6        t[i]= t[k];
7        t[k] = tmp;    }  }
/*@ requires 0<=l && l<t.length
 @ ensures  (l<=\result) && (\result<t.length)
 @       && (\forall int k; l<=k && k<t.length;t[\result]<=t[k]) @*/
1  static int findMin(int[] t,int l) {
2    int idx = l;
3    for (int j = l+1; j < t.length;j++)
4        if (t[idx]>t[j])
5           idx = j;
6    return idx; }
\end{verbatim}}

\subsection{Modular verification and ``element constraint''}
Assume that function
\verb+findMin+ has been verified for arbitrary integers. When
encountering a call to \verb+findMin+, CPBPV  first checks
if its precondition is entailed by the constraint store, which
requires a consistency check of the constraint store with respect to
the negation of the precondition. Then CPBPV  replaces
the call by the post-condition where the formal parameters are
replaced by the actual variables. In particular, for the first
iteration of the loop and an array length of 40, CPBPV 
generates the constraint
\[
0 \leq k^0 < 40 \; \wedge \; t^0[k^0] \leq t^0[0] \; \wedge \; \ldots \; \wedge \; t^0[k^0] \leq t^0[39].
\]
This constraint is interesting, since it features {\tt element}
constraint \cite{VanHentenryck89}, i.e., the ability of indexing
arrays with expressions containing variables. Indeed, $k^0$ is a
variable and a constraint like $t^0[k^0] \leq t^0[0]$ indexes the
array $t^0$ of variables using $k^0$. 
The {\tt element} constraint is
an important functionality of constraint programming, not only because
of its ubiquity in practice but also because it highlights the kind of
symbolic processing and filtering allowed by this technology. Note
also that the subsequent assignments also create {\tt element}
constraints.

\subsection{Comparative results }
The modular verification of the selection sort explores only a single
path, is independent of the integer representation, and takes less
than $0.01s$ for arrays of size 40. The bottleneck in verifying
selection sort is the validation of function \verb+findMin+, which
requires the exploration of many paths. However the complete
validation of selection sort takes less than 4 seconds for an array of
length 6. Once again, this should be contrasted with the
model-checking approach of Eureka \cite{ABM07}. On a version of
selection sort where all variables are assigned specific values
(contrary to our verification which makes no assumptions on the
inputs), Eureka takes 104 seconds on a faster machine. Reference
\cite{ABM07} also reports that CBMC takes 432.6 seconds, that BLAST
cannot solve this problem, and that SATABS \cite{CKS05} only verifies
the program for an array with 2 elements.


\end{document}